\pdfoutput=1
\documentclass{ptephy}

\preprintnumber{XXXX-XXXX} 

\usepackage{graphicx,amsmath,amssymb}
\usepackage{color}
\usepackage{cite} 
\usepackage{subfig} 
\usepackage[normalem]{ulem}

\newcommand{\rout}{\bgroup \color{red} \ULdepth=-.5ex \ULset}
\newcommand\bout{\bgroup \color{blue} \ULdepth=-.5ex \ULset}

\begin{document}

\title{Magnetism and rotation in relativistic field theory}

\author{\name{Kazuya~Mameda}{} and \name{Arata~Yamamoto}{}}
\address{\affil{}{Department of Physics, The University of Tokyo, Tokyo 113-0033, Japan}
}

\begin{abstract}%
We investigate the analogy between magnetism and rotation in relativistic theory.
In nonrelativistic theory, the exact correspondence between magnetism and rotation is established in the presence of an external trapping potential.
Based on this, we analyze relativistic rotation under external trapping potentials.
A Landau-like quantization is obtained by considering an energy-dependent potential.
\end{abstract}

\subjectindex{B30, E76}

\maketitle


\section{Introduction}
In quantum physics, a response to magnetism is one of the most significant subjects.
One ubiquitous phenomenon induced by external magnetic backgrounds is the Landau quantization.
In condensed matter systems, for example, the quantum Hall effect is essentially described by the Landau levels~\cite{Girvin:1987}.
In relativistic theory, such as quantum chromodynamics, an external magnetic field enhances a fermion-antifermion condensate and generates a dynamical fermion mass, which is called the magnetic catalysis~\cite{Gusynin:1994,Gusynin:1995}.
The magnetic catalysis is also important in condensed matter systems with relativistic dispersion relations, such as graphenes and Dirac semimetals~\cite{Miransky:2015ava}.

Rotation has similar effects to magnetism,
and so they share many common phenomena:
The quantum Hall effect is also induced by rotation instead of an external magnetic field~\cite{Viefers}.
A quantum vortex is generated by applying a magnetic field to the Bose-Einstein condensate or by rotating it~\cite{Fetter,Tsubota}.
The chiral magnetic effect~\cite{KharCME,FukuCME} is analogous to the chiral vortical effect~\cite{KharCVE,SonCVE}, which can be generated by the vorticity in rotating quark-gluon plasma~\cite{Kharzeev:2015znc,Jiang:2016woz,Deng:2016gyh}.
Note that, since rotational effects are independent of the electric charge, these rotational phenomena emerge even for neutral particles while magnetic phenomena arise only for charged particles.

Indeed, in nonrelativistic theory, the exact correspondence between magnetism and rotation is known~\cite{Coisson:1973,Sivardiere:1983}.
By rotating an atomic gas in a harmonic trap very rapidly, we can experimentally realize this correspondence~\cite{Wilkin,Cooper,Schweikhard}.
If this correspondence is achieved, the Landau quantization and other magnetic phenomena, such as quantum Hall effect, are expected to be observable in rotating media.
On the other hand, the relation between magnetism and rotation is not clear in relativistic theory.
Since relativistic theory is more fundamental than nonrelativistic theory, which is derived by taking the nonrelativistic limit of relativistic theory, the nonrelativistic correspondence must be potentially included in relativistic theory.
If we can find some correspondence, we can advocate novel phenomena in relativistic rotating matters, which are analogous to magnetic phenomena.

In this paper, we discuss the analogy between magnetism and rotation in relativistic theory.
As reviewed in the next section, the exact correspondence in nonrelativistic theory is obtained only after applying an external trapping potential.
Based on this, we analyze rotating systems with external trapping potentials in relativistic theory and discuss the quantization of the energy spectrum.

\section{Nonrelativistic theory}
In this section, we review the established relation between magnetism and rotation in nonrelativistic theory.

As the simplest example, let us illustrate the correspondence in classical mechanics.
A charged particle in a magnetic field receives the Lorentz force
\begin{equation}\label{eq:Fmag}
{\boldsymbol F}=e{\boldsymbol v}\times{\boldsymbol B}.
\end{equation}
On the other hand, a rotating particle feels two apparent forces,
\begin{equation}\label{eq:Frot}
{\boldsymbol F}= 2m{\boldsymbol v}\times{\boldsymbol\Omega}-m{\boldsymbol \Omega}\times({\boldsymbol \Omega}\times{\boldsymbol x}),
\end{equation}  
i.e., the Coriolis force and the centrifugal force, respectively.
The Coriolis force is mathematically equivalent to the Lorentz force under the replacement
$e{\boldsymbol B} \leftrightarrow 2m{\boldsymbol \Omega}$.
Except for the centrifugal force, therefore, we see the correspondence between magnetism and rotation.
In other words, the crucial point for the correspondence is how to eliminate the centrifugal force.

Also in nonrelativistic quantum mechanics, it can be shown that the centrifugal force is the difference between magnetism and rotation.
Suppose a magnetic field and an angular velocity along the $z$-axis, ${\boldsymbol B} = B\hat z $ and ${\boldsymbol \Omega}= \Omega \hat z$.
The Hamiltonian of a charged particle in a magnetic field is given by
\begin{equation}\label{eq:Hmag}
\begin{split}
H = \frac{1}{2m}({\boldsymbol p}-e{\boldsymbol A})^2  = \frac{1}{2m} ({\boldsymbol p} - eB \hat z \times {\boldsymbol x}/2 )^2,
\end{split}
\end{equation}
with the symmetric gauge ${\boldsymbol A}=(-By/2, Bx/2,0)$.
The energy spectrum is quantized as
\begin{equation}\label{eq:magNRDR}
E = \frac{eB}{2m} (2n+1) + \frac{p_z^2}{2m},
\end{equation}
which is called the Landau quantization.
On the other hand, the Hamiltonian for a particle in a rotating frame reads
\begin{equation}\label{eq:Hrot}
\begin{split}
H &= \frac{\boldsymbol p^2}{2 m} - \Omega \hat z \cdot ({\boldsymbol x} \times {\boldsymbol p}) 
= \frac{1}{2m} ({\boldsymbol p} - m\Omega \hat z \times {\boldsymbol x} )^2 - \frac{1} {2} m \Omega^2 {(x^2 + y^2)} .
\end{split}
\end{equation}
Also here, the correspondence between the kinetic terms in Eqs.~\eqref{eq:Hmag} and \eqref{eq:Hrot} is confirmed.
However, the second term in Eq.~\eqref{eq:Hrot} is missing in Eq.~\eqref{eq:Hmag}.
As in classical mechanics, this superfluous term is related to the centrifugal force.
Indeed, we obtain the centrifugal force in Eq.~\eqref{eq:Frot} from
\begin{equation}
\nabla \left[ -\frac{1}{2} m \Omega^2(x^2 + y^2) \right] = m{\boldsymbol \Omega}\times({\boldsymbol \Omega}\times{\boldsymbol x}).
\end{equation}
This centrifugal force potential creates the different physical situation from that of magnetism with respect to homogeneity.
The presence of the centrifugal force potential gives the inhomogeneity in rotating frames, while systems under external magnetic fields are homogeneous.
If we can eliminate the centrifugal force potential by applying the external trapping potential
\begin{equation}
 V({\boldsymbol x}) = \frac{1}{2} m \Omega^2(x^2 + y^2)
\end{equation}
by hand, we can observe the Landau quantized energy spectrum~\cite{Cooper}
\begin{equation}\label{eq:rotNRDR}
E = \Omega (2n+1) + \frac{p_z^2}{2m}.
\end{equation}
Therefore, when the centrifugal force potential and the external trapping potential cancel out, magnetism and rotation is equivalent with the correspondence
\begin{equation}\label{eq:nonrelamagrot}
 eB \leftrightarrow 2m\Omega.
\end{equation}

\section{Relativistic theory}

We consider the relativistic scalar field theory in the cylindrical coordinate $x^\mu = (t,r,\theta,z)$.
The solutions for the following equations are written as
\begin{equation}\label{eq:scalar}
 \phi(x) = e^{-i\varepsilon t +ip_z z + i\ell\theta} \Phi(r),
\end{equation}
where the $z$-axis is the direction of $B$ and $\Omega$.

In the presence of a magnetic field, the Klein-Gordon equation is
\begin{equation}\label{eq:magKG0}
\begin{split}
\bigg[\partial_t^2 - \partial_r^2 -\frac{1}{r} \partial_r - \frac{1}{r^2} \partial^2_\theta - \partial_z^2 + m^2 
+ieB\partial_\theta + \frac{1}{4}e^2B^2 r^2 \bigg] \phi(x) = 0 .
\end{split}
\end{equation}
We take $eB>0$.
Substituting Eq.~\eqref{eq:scalar}, we obtain
\begin{equation}\label{eq:magKG}
\begin{split}
\bigg[ \varepsilon^2 + \partial_r^2 + \frac{1}{r} \partial_r - \frac{\ell^2}{r^2} - p_z^2 - m^2 
+ eB\ell - \frac{1}{4} e^2B^2 r^2 \bigg] \Phi(r) = 0 .
\end{split}
\end{equation}
The solution of this equation is known as the Landau wave function
\begin{equation}\label{eq:magsol}
\Phi(r) =  r^{\ell} e^{-eBr^2/4} L_n^{\ell} (eBr^2/2),
\end{equation}
where $L^k_n(x)$ is an associated Laguerre polynomial. 
The corresponding energy eigenvalue, i.e., the Landau energy level, is given by
\begin{equation}\label{eq:magDR}
\varepsilon = \pm\sqrt{eB(2n+1) + p_z^2 + m^2}.
\end{equation}
The Landau levels depends only on the radial quantum number $n$ and not on the azimuthal quantum number $\ell$, and so the Landau states with different $\ell$ are energetically degenerate.
The normalizability of the wave function \eqref{eq:magsol} restricts the range of $\ell$ from $-n$ to $\infty$.
We should mention a necessary condition for the Landau quantization.
In general the quantization for harmonic oscillators demands that the typical scale of the trapping potential should be much smaller than the system size. 
To obtain the wave function \eqref{eq:magsol} and the energy level \eqref{eq:magDR}, we thus need a system with a radius that is large enough compared with the magnetic length:
\begin{equation}
1/\sqrt{eB} \ll R.
\label{eq:magcondition}
\end{equation}
In the case that Eq.~\eqref{eq:magcondition} is not satisfied, the wave function is not localized enough and thus we are to consider a boundary effect.

In relativistic theory, rotation is described as spacetime geometry through the corresponding metric tensor.
In a curved spacetime with $g_{\mu\nu}$, the Klein-Gordon equation is
\begin{equation}\label{eq:rotKG0}
\left [ \frac{1}{\sqrt{-\det g}} \partial_\mu \sqrt{-\det g} g^{\mu\nu} \partial_\nu + m^2 + V(r) \right ] \phi(x) = 0,
\end{equation}
where $V(r)$ is an external potential.
In the rotating frame, the metric tensor is
\begin{equation}\label{eq:cymetric}
g_{\mu\nu} =
\begin{pmatrix}
1-r^2\Omega^2 & 0 & -r^2\Omega & 0 \\
0 & -1 & 0 & 0 \\
-r^2\Omega & 0 & -r^2 & 0 \\
0 & 0 & 0 & -1 \\
\end{pmatrix}
.
\end{equation}
We take $\Omega>0$.
Substituting Eqs.~\eqref{eq:scalar} and \eqref{eq:cymetric} to Eq.~\eqref{eq:rotKG0}, we obtain the equation
\begin{equation}\label{eq:rotKG}
 \bigg[ (\varepsilon + \ell \Omega)^2 + \partial_r^2 + \frac{1}{r} \partial_r - \frac{\ell^2}{r^2} - p_z^2 - m^2 - V(r) \bigg] \Phi(r) = 0.
\end{equation}
This equation can be also derived by replacing the energy $\varepsilon$ on the flat Minkowski space with $\varepsilon + \ell \Omega$.
Because of the relativistic causality, we must consider the rotating cylinder with a finite radius $R \leq 1/\Omega$; otherwise, the frame moves faster than light.
In a rotating frame with $R>1/\Omega$, particle spectrum is known to be pathological~\cite{Letaw:1979wy,Vilenkin:1980zv,Davies:1996ks,Duffy:2002ss,Ambrus:2014uqa}. 

In order to compare rotation with magnetism, we analyze the energy spectrum of Eq.~\eqref{eq:rotKG} in three cases of the external potential $V(r)$.

(I)
Based on the discussion of the nonrelativistic case, we consider a harmonic trapping potential with the constant parameter $\alpha>0$:
\begin{equation}\label{eq:V1}
 V(r)= \alpha^2\Omega^2 r^2 .
\end{equation}
In this case, Eq.~\eqref{eq:rotKG} can be easily solved, if we impose the similar condition to Eq.~\eqref{eq:magcondition}:
\begin{equation}
1/\sqrt{\alpha\Omega} \ll R \leq 1/\Omega.
\label{eq:rot1condition}
\end{equation}
As long as the first inequality is satisfied, the wave function is localized and exponentially suppressed at the boundary.
Therefore, similarly to the magnetic Landau quantization, the boundary effect can be ignored.
The second inequality is imposed for keeping causality as mentioned above.
The solution of Eq.~\eqref{eq:rotKG} is
\begin{equation}\label{eq:trapsol1}
\Phi(r) = r^{\ell} e^{-\alpha\Omega r^2/2} L_n^{\ell} (\alpha\Omega r^2/2),
\end{equation}
with the energy spectrum
\begin{equation}\label{eq:trapDR1}
\varepsilon = -\ell\Omega \pm \sqrt{2\alpha\Omega(2n+\ell+1) + p_z^2 + m^2}.
\end{equation}
The energy spectrum depends on both $n$ and $\ell$.
This is quite different from magnetism.

For the nonrelativistic expansion with $\alpha = m$, we explicitly show the speed of light $c$: $m= mc^2$, $\Omega=\Omega/c$ and $p_z=p_zc$.
The nonrelativistic energy is then given as
\begin{equation}
\begin{split}
E =& \ \varepsilon - mc^2 \\
  =& \ \frac{p_z^2}{2m} + \frac{\Omega}{c}(2n+1)
  - \frac{p_z^4}{4m^3c^2} - \frac{p_z^2\Omega}{m^2c^3}(2n+\ell+1) \\
  &- \frac{\Omega^2}{mc^4}(2n+\ell+1)^2 + \frac{3p_z^6}{8m^5c^4} + O(1/c^5)
\end{split}
\end{equation}
The first two terms correspond to the nonrelativistic Landau quantization \eqref{eq:rotNRDR}.
Other terms are higher-order relativistic corrections.
Although the terms proportional to $(2n+\ell+1)$ and $(2n+\ell+1)^2$ seem to interrupt the realization of the nonrelativistic Landau quantization, such terms can actually be negligible because Eq.~\eqref{eq:rot1condition} implies the inequality $\Omega \ll mc^3$.

(II)
To get a similar energy spectrum to the Landau quantization, we need an energy-dependent potential
\begin{equation}\label{eq:V2}
 V(r) = \ell^2\Omega^2 + \varepsilon^2 \Omega^2 r^2.
\end{equation}
The equation becomes
\begin{equation}\label{eq:trapKG2}
\begin{split}
\bigg[ \varepsilon^2 + \partial_r^2 + \frac{1}{r} \partial_r - \frac{\ell^2}{r^2} - p_z^2 - m^2 
+2 \varepsilon \Omega \ell - \varepsilon^2 \Omega^2 r^2 \bigg] \Phi(r) = 0,
\end{split}
\end{equation}
which is the same form as Eq.~\eqref{eq:magKG} with the correspondence
\begin{equation}\label{eq:relamagrot}
eB \leftrightarrow 2\varepsilon\Omega.
\end{equation}
Note that $\varepsilon$ can be positive and negative unlike $eB$ and $m\Omega$. (Here, we take $\Omega>0$.)
Since the transformation $\varepsilon \to - \varepsilon$ is equivalent to the parity transformation $\ell \to -\ell$ in Eq.~\eqref{eq:trapKG2}, the solution with $\varepsilon <0$ is obtained by the replacement $\ell \to -\ell$ in Eq.~\eqref{eq:trapKG2}.
We will write the solution as a function of $|\varepsilon|$ and take $+\ell$ ($-\ell$) for $\varepsilon >0$ ($\varepsilon <0$).
The correspondence \eqref{eq:relamagrot} leads to the following solution:
\begin{equation}\label{eq:trapsol}
\Phi(r) =  r^{\pm\ell} e^{-|\varepsilon|\Omega r^2/2} L_n^{\pm\ell} (|\varepsilon|\Omega r^2).
\end{equation}
The energy spectrum is
\begin{equation}\label{eq:trapDR2}
|\varepsilon| = \Omega(2n+1) + \sqrt{ \Omega^2  (2n + 1)^2  + p_z^2 + m^2},
\end{equation}
which has positive-energy and negative-energy modes.
In the sense that the energy depends only on $n$ and not on $\ell$, this is a ``Landau-like'' quantization.
At $p_z=m=0$, the energy is proportional to $\Omega$, $|\varepsilon| = 2\Omega(2n+1)$. 
We mention that such a spectrum is realized only for $n\gg 1$ because this quantization demands
\begin{equation}
1/\sqrt{|\varepsilon|\Omega} \ll R \leq 1/\Omega.
\label{eq:rot2condition}
\end{equation}
In the nonrelativistic limit, the external potential \eqref{eq:V2} is reduced to Eq.~\eqref{eq:V1} and the energy reproduces Eq.~\eqref{eq:rotNRDR}.

(III)
If we drop the $\ell$-dependent potential in Eq.~\eqref{eq:V2}, and apply only the trapping potential
\begin{equation}
 V(r) = \varepsilon^2 \Omega^2 r^2,
\end{equation}
then the solution is the same as Eq.~\eqref{eq:trapsol}, but the energy spectrum is modified as
\begin{equation}\label{eq:trapDR3}
|\varepsilon| = \Omega(2n+1)\pm \sqrt{ \Omega^2  \left[( 2n + 1)^2 - \ell^2 \right]  + p_z^2 + m^2}.
\end{equation}
Again we consider that Eq.~\eqref{eq:rot2condition} is satisfied.
There are four branches of the dispersion relation.
Such a multi-branch structure is explained by the Klein-Gordon equation \eqref{eq:rotKG}.
For $V(r)=0$, since the energy appears only in the term $(\varepsilon+\ell\Omega)^2$, there are two branches corresponding to the sign of $\varepsilon+\ell\Omega$, as shown in Eq.~\eqref{eq:freeDR}.
On the other hand, since Eq.~\eqref{eq:trapKG2} includes not only $(\varepsilon+\ell\Omega)^2$ but also $\varepsilon^2$, there are four branches corresponding the signs of $\varepsilon+\ell\Omega$ and $\varepsilon$.
This is related to whether to regard $\phi(x)$ as a particle or an antiparticle.
For example, even if $\phi(x)$ in the rotating frame is considered as a particle (i.e. $\varepsilon>0$), there are both cases that $\phi(x)$ in the inertial frame is regarded as a particle (i.e. $\varepsilon+\ell\Omega>0$) and as an antiparticle (i.e. $\varepsilon+\ell\Omega<0$)~\cite{Letaw:1979wy}.
The $\ell$-dependence in Eq.~\eqref{eq:trapDR3} plays a similar role to the spin-dependence of a charged particle under magnetic fields,
\begin{equation}\label{eq:NOinsta}
\varepsilon = \pm\sqrt{eB(2n+1-2s)+p_z^2+m^2}.
\end{equation}
Thus, large-$\ell$ modes are tachyonic like the Nielsen-Olesen unstable modes~\cite{Nielsen:1979xu}.
We should mention that the difference between these two instabilities.
The Nielsen-Olesen instability results from only the lower modes with $n\lesssim s -1/2$.
On the other hand our instability requires not only $n\lesssim (|\ell|-1)/2$ but also $n\gg 1$, which is led by Eq.~\eqref{eq:rot2condition}. Therefore in Eq.~\eqref{eq:trapDR3} only higher modes generate the instability.
We also note that both the multi-branch dispersion and the instability result from the superfluous term of rotation, i.e. $\ell^2\Omega^2$, as shown in Eqs.~\eqref{eq:trapDR2} and \eqref{eq:trapDR3}.

\section{Scalar condensate}

One important property of magnetism is the catalyzing effect on a condensate, which is an order parameter of spontaneous symmetry breaking. 
Let us briefly discuss rotational effects on a scalar condensate.
The particle spectrum in a rotating frame is trivially obtained by the coordinate transformation from the Minkowski space.
This transformation does not change the condensate in each vacuum, i.e., $\langle 0_\text{Min}| \phi_\text{Min}| 0_\text{Min} \rangle = \langle 0_\text{Rot}| \phi_\text{Rot}| 0_\text{Rot} \rangle$. 
However, it can change the condensate observed in the different vacuum, i.e., $\langle 0_\text{Min}| \phi_\text{Min}| 0_\text{Min} \rangle \neq \langle 0_\text{Min}| \phi_\text{Rot}| 0_\text{Min} \rangle$.

\begin{figure}[!h]
\centering\includegraphics[width=3.5in]{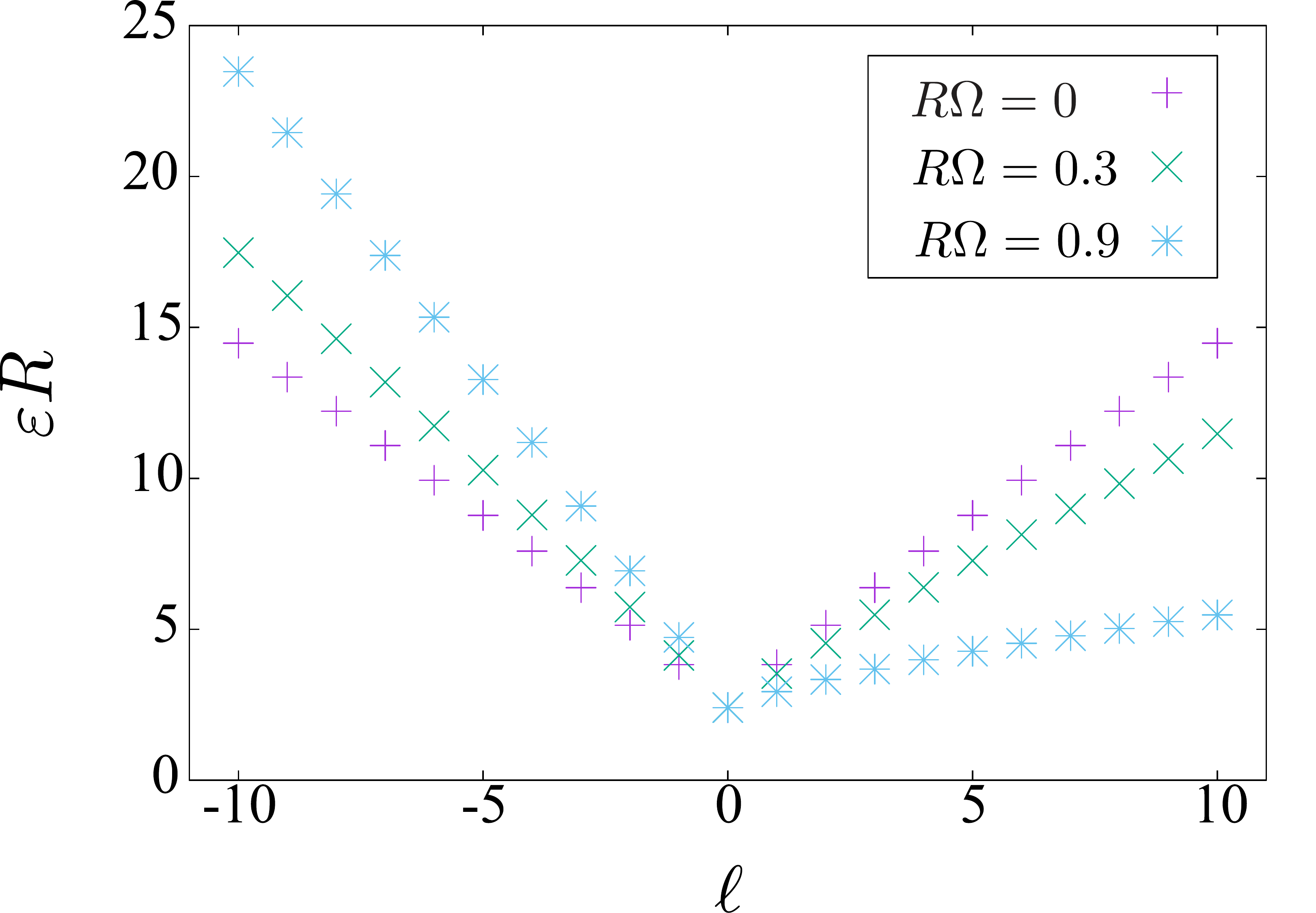}
\caption{
Positive sign spectrum in Eq.~\eqref{eq:freeDR} with $p_z=m=0$ and $p_\perp=\xi_{\ell 1}/R$.
The $\ell=0$ mode always corresponds to the ground state.
In rotating frames ($\Omega\neq 0$), the positive $\ell$ modes are more energetically-favorable than the $\ell$ modes.
}
\label{fig:bessel}
\end{figure}

We first consider the energy spectrum of noninteracting scalar theory without external potentials.
The solution of Eq.~\eqref{eq:rotKG} for $V(r)=0$ is
\begin{equation}\label{eq:freesol}
\Phi(r) = J_\ell (p_\perp r)
\end{equation}
with the energy 
\begin{equation}\label{eq:freeDR}
\varepsilon = - \ell\Omega \pm \sqrt{p_\perp^2 + p_z^2 + m^2}.
\end{equation}
Since the angular momentum $\ell$ can be arbitrary integer, this spectrum seems unbounded.
This problem comes from the invalid assumption that the system size is infinitely large.
In a rotating system with $R \leq 1/\Omega$,
the perpendicular momentum $p_\perp$ should be discretized.
For example if we impose the Dirichlet boundary condition at $r=R$, the discrete momentum is given by
\begin{equation}\label{eq:pperp}
p_\perp = \xi_{\ell k}/R,
\end{equation}
where $\xi_{\ell k}$ is the $k$th zero of $J_\ell(z)$.
We note that the Bessel zeros satisfies $\xi_{\ell 1}>|\ell|$~\cite{Watson}.
As long as causality is kept (i.e. $R\Omega\leq 1$), hence, the positive-sign spectrum in Eq.~\eqref{eq:freeDR} has a lower-bound:
\begin{equation}\label{eq:IR}
\varepsilon \geq -\ell\Omega + \xi_{\ell 1}/R > 0.
\end{equation}
In Fig.~\ref{fig:bessel}, we plot the lowest spectrum  for $p_z=m=0$, i.e. $\varepsilon = -\ell\Omega + \xi_{\ell 1}/R$. 
The ground state is always $l=0$ even for $\Omega > 0$.

If the system has a self-interaction and a chemical potential, the scalar condensate can be nonzero. 
For $V(r) = 0$, the change of the condensate stems from the shift of the energy dispersion $\varepsilon\to\varepsilon+\ell\Omega$. 
This shift is regarded as an effective chemical potential $\mu_\ell \equiv \ell \Omega$~\cite{Chen:2015hfc,Jiang:2016wvv}.
The condensate component $l=0$ is not directly affected by this effective chemical potential while the normal component $l\ne 0$ is affected.
The condensate changes through the self-interaction with the normal component.
In addition, the particle number density and the condensate become inhomogeneous.
A na\"{i}ve expectation is that the number density increases and the condensate decreases in large $r$ due to a centrifugal force.

If an external potential is introduced in the rotating system, the effect on the condensate would be more nontrivial.
For example, the ground state could change from $\ell = 0$ to $\ell \neq 0$. 
(For the evaluation of the true ground state, we have to precisely take account of a boundary condition, which we have simplified in the previous section.)
Moreover, we expect some nontrivial generation of the condensate, namely, the `rotational catalysis', which is analogous to the magnetic catalysis.
Since rotation couples to all kinds of fields, even to charge-neutral fields, we can universally expect it in various systems.
Such a novel phenomenon induced by rotation is a quite fascinating work based on the discussion in this paper. 

\section{Summary}
We investigated the correspondence between magnetism and rotation in relativistic theory.
We discussed the Landau quantization by nonrelativistic rotation and evaluated the relativistic correction to it.
Considering the energy-dependent trapping potential, we found the Landau-like quantization by relativistic rotation.

In this paper, we discussed the scalar field theory for simplicity.
The analysis can be extended to higher spin fields.
For example, a rotating spin-$1/2$ fermion is described by the Dirac equation in a rotating frame.
The Dirac equation in a rotating frame has the spin-rotation coupling term, which corresponds to the Zeeman term in magnetism, although it depends on the choice of vierbein~\cite{Iyer:1982ah,Hehl:1990nf}.
The momentum quantization for the fermion is not as trivial as Eq.~\eqref{eq:pperp} because of the spin-rotation coupling.
Imposing a boundary condition in terms of a current conservation, we obtain a quantized momentum for the fermion~\cite{Ambrus:2015lfr,Ebihara:2016fwa}.
In this case, the $\ell$-dependence of the lowest energy mode is slightly changed (all the points in Fig.~\ref{fig:bessel} move to left by 1/2), but we can still confirm that the dispersion is lower-bounded.
Therefore, the same discussion would be possible for the rotational effects on the fermionic energy spectrum and condensate.

\section*{Acknowledgements}
The authors thank Kenji Fukushima, Sanjin Benic, Tomoya Hayata and Patrick Copinger for useful discussions.
K.~M.~is supported by JSPS Research Fellowships for Young Scientists.
A.~Y.~is supported by JSPS KAKENHI Grants Number 15K17624.


%

\end{document}